\documentclass[10pt,aps,prb,superscriptaddress,reprint]{revtex4-1}
\usepackage{graphicx}
\usepackage{amsmath}
\usepackage[english]{babel}
\usepackage{array}
\usepackage{amssymb, bm}
\usepackage{multirow}
\usepackage{epstopdf}

\newcommand{\ds}{\displaystyle}

\begin{document}
\title{Crossover between standard and inverse spin-valve effect in atomically thin superconductor/half-metal structures}

\author{Zh. Devizorova}
\affiliation{Moscow Institute of Physics and Technology, 141700 Dolgoprudny, Russia}
\affiliation{Kotelnikov Institute of Radio-engineering and Electronics RAS, 125009 Moscow, Russia}
\affiliation{Institute for Physics of Microstructures, Russian Academy of Sciences, 603950 Nizhny Novgorod, GSP-105, Russia}
\author{S. Mironov}
\affiliation{Institute for Physics of Microstructures, Russian Academy of Sciences, 603950 Nizhny Novgorod, GSP-105, Russia}

\begin{abstract}
Spin-singlet Cooper pairs consisting of two electrons with opposite spins cannot directly penetrate from a superconductor to a half-metal (fully spin polarized ferromagnets) which blocks the superconducting proximity effect between these materials. In this paper we demonstrate that, nevertheless, two half-metallic layers electrically coupled to the superconducting film substantially affect its critical temperature and produce the spin valve effect. Within the tight-binding model for the atomically thin multilayered spin valves we show that depending on the details of the electron energy spectra in half-metals the critical temperature as a function of the angle between the spin quantization axes in half-metals can be either monotonically increasing or decreasing. This finding highlights the crucial role of the band structure details in the proximity effect with half-metals which cannot be adequately treated in the quasiclassical theories.
\end{abstract}

\maketitle

\section{Introduction}

The phenomena originating from the exchange of electrons between superconductors and multilayered ferromagnets have the great potential for the application in superconducting spintronics \cite{Linder_rev, Eschrig_rev} since they provide an efficient tool for the control of the charge and spin transport by changing the magnetic state of the ferromagnet. The basic control element (so-called superconducting spin valve) consisting of a thin superconducting (S) film and two ferromagnets (F) performs as the superconducting analog of  transistor controlled by an external magnetic field \cite{Oh_APL, Tagirov_PRL, Buzdin_EPL99}. The critical temperature $T_c$ of such structure strongly depends on the angle $\theta$ between the magnetic moments in the ferromagnets. Thus, fixing the system temperature between the minimum $T_c^{min}$ and the maximum $T_c^{max}$ of the critical temperature and changing the mutual orientation of the magnetic moment of the F layers by the external magnetic field one can significantly vary the resistivity of the spin valve switching it from the normal to the superconducting state (spin-valve effect). 

The physics behind the strong dependence $T_c(\theta)$ is related to the superconducting proximity effect \cite{Buzdin_RMP, BVE_RMP}. The exchange field in the ferromagnet destroys the Cooper pairs and change their spin structure. This results in the peculiar damped oscillatory behavior of the Cooper pair wave function inside the F layers and damping of the superconductor critical temperature. If the thickness of the ferromagnets is small compared to the coherence length $\xi_f$ characterizing the oscillations period then the critical temperature is determined simply by the average exchange field and, therefore, the function $T_c(\theta)$ is monotonically increasing and $T_c(\pi)>T_c(0)$ (so-called standard spin-valve effect). \cite{Tagirov_PRL, Buzdin_EPL99, Baladie_PRB, You_PRB, Buzdin_EPL03, Tollis_PRB, Bozovic_EPL, Halterman_PRB, Linder_PRB, Lofwander_PRB} For the F layers with the thickness $\sim \xi_f$ the interference phenomena coming from the oscillations of the wave function make $T_c(\pi)<T_c(0)$ for the certain range of parameters (inverse spin-valve effect). \cite{Buzdin_RMP, Fominov_JETPL, Mironov_PRB14}  Moreover, the non-collinearity of the magnetic moment orientation in the F layers produces the long-range spin-triplet correlations \cite{Bergeret_PRL} which form an additional channel for the Cooper pair leakage from the superconductor and, thus, increase the damping of $T_c$. As a result, for the certain parameters the minimum of $T_c$ corresponds to $\theta\not= 0, \pi$ (so-called triplet spin-valve effect). \cite{Fominov_JETPL, Wu_PRB, Mironov_PRB14, Devizorova_PRB}

Experimentally the spin-valve effect was observed in a wide class of F$_1$/S/F$_2$ \cite{Nowak_PRB, Kinsey, Gu_PRL, Potenza_PRB, Westerholt_PRL, Moraru_PRL, Moraru_PRB, Kim, Luo_EPL, Zhu_PRL, Rusanov_PRB, Aarts, Steiner_PRB, Singh_APL, Singh_PRB, Leksin_JETPL,Floksta_PRB}  and S/F$_1$/F$_2$ \cite{Nowak_PRB, Nowak_SST, Leksin_PRL11,  Leksin_PRL12, Leksin_PRB, Zdravkov_PRB13}  structures. The magnitude of the effect appears to be very sensitive to the choice of ferromagnetic materials. Indeed, the typical scale of the Cooper pair wave function decay in ferromagnets tends to decrease with the increase in the exchange field. Therefore, the vast majority of spin valves are based on the ferromagnetic alloys (e.g., CuNi or PdFe) with small exchange field compared to the Fermi energy. However, such structures are hardly applicable for the devices of superconducting spintronics since the variation of their critical temperature $\Delta T_c=T_c^{max}-T_c^{min}$  does not exceed several percents. 

Recently it was demonstrated that the magnitude of the spin-valve effect can be significantly increased \cite{Singh_PRX} provided one of the ferromagnetic layers is made of half-metal (HM), the material with the exchange field comparable to the Fermi energy (e.g., Co, CrO$_2$).\cite{Pickett_PT, Coey_JAP} The full spin polarization of electrons in half-metals make them extremely promising materials for the superconducting spintronics. However, the quantitative theoretical description of the proximity effect in S/HM structures appears to be challenging due to the breakdown of the quasiclassical approximation which requires small exchange fields and energy shifts between electron energy bands in different layers compared to the Fermi energy.  Despite several attempts to develop quasiclassical theory of the superconducting proximity effect with half-metals \cite{Eschrig_NJP, Moor_PRB, Mironov_PRB15, EschrigLinder} the quantitative quasiclassical description of such materials is still lacking. An alternative numerical solution of the Bogoliubov-de Gennes equations supports the experimentally observed increase of the spin-valve effect in HM-based systems \cite{Halt1, Halt2}. At the same time, the exact analytical solutions of the Gor'kov equations for the atomically thin S/F/HM heterostructures beyond the quasiclassical approximation additionally demonstrate the strong sensitivity of the spin-valve effect to the details of the electronic energy band structure inside each of the layers \cite{Devizorova_PRB}. Specifically, depending on the relative shift in the electron bands in different layers the dependence $T_c(\theta)$ approaches its minimum at $\theta=0$ or $\theta=\pi$ which corresponds to the standard or inverse spin-valve effect.  Thus, the adequate theoretical description of the spin-valve effect in the superconducting hybrids containing half-metals requires the accurate account of the band structure effects which cannot be done within the quasiclassical approaches.   

Since half-metals can host only spin-1 triplet superconducting correlations their direct contact with singlet s-wave superconductor should not give rise to the proximity effect. As a result, the conventional design of the HM-based spin valve contains an additional ferromagnetic layer with the small exchange field or other type of spin-active interface. Such additional layer modifies the spin structure of the Cooper pairs and generates spin-triplet correlations which can penetrate the half-metallic layer \cite{Singh_PRX, Robinson}. Interestingly, even if the superconductor is placed between two half-metals its critical temperature depends on the mutual orientation of the spin quantization axes in the HM layers due to non-local effects\cite{Montiel_EPL}. The exact solution of the Gor'kov equations for the atomically thin HM/S/HM structure with two identical HM layers predicts both standard and inverse spin-valve effect depending on the shift between the bottom of the energy bands in each half-metal and the one in the S layer. Remarkably, the situation $T_c(0)>T_c(\pi)$ was found only for the very specific case when the electron spectrum in one of two HM layers is hole-like. 

In the present paper we analyze the possible types of the spin-valve effect in atomically thin HM$_1$/S/HM$_2$ and  S/HM$_1$/HM$_2$ structures. We assume the electron-like spectrum in each of the HM layers and take into account the dispersion of the only occupied energy band (in contrast to Ref.~\onlinecite{Montiel_EPL}). We find that the details of the electron band structure in the half-metallic layers have the major influence on the type of the spin-valve effect. Specifically, the relative shift between these bands in two HM layers can lead to the inversion of the spin-valve effect [which corresponds to the situation $T_c(0)>T_c(\pi)$] even without the sign change in the electron effective mass. Our finding shows that combining different half-metals in the spin-valve one can tune the dependence $T_c(\theta)$ making it either increasing or decreasing.

The paper is organized as follows. In Sec. \ref{Sec:SHM1HM2} we consider the S/HM$_1$/HM$_2$ structure in which the occupied spin band in HM$_1$ or HM$_2$ layer is shifted with respect to the electron energy band in the S layer and analyze the effect of this shift on the behavior of the critical temperature. The Sec. \ref{Sec:HM1SHM2} is devoted to the spin-valve effect in HM$_1$/S/HM$_2$ structure. In Sec. \ref{Sec:Conclusion} we summarize our results.

\section{Spin valve effect in S/HM1/HM2 structure} \label{Sec:SHM1HM2}

\begin{figure}[t!]
\includegraphics[width=0.3\textwidth]{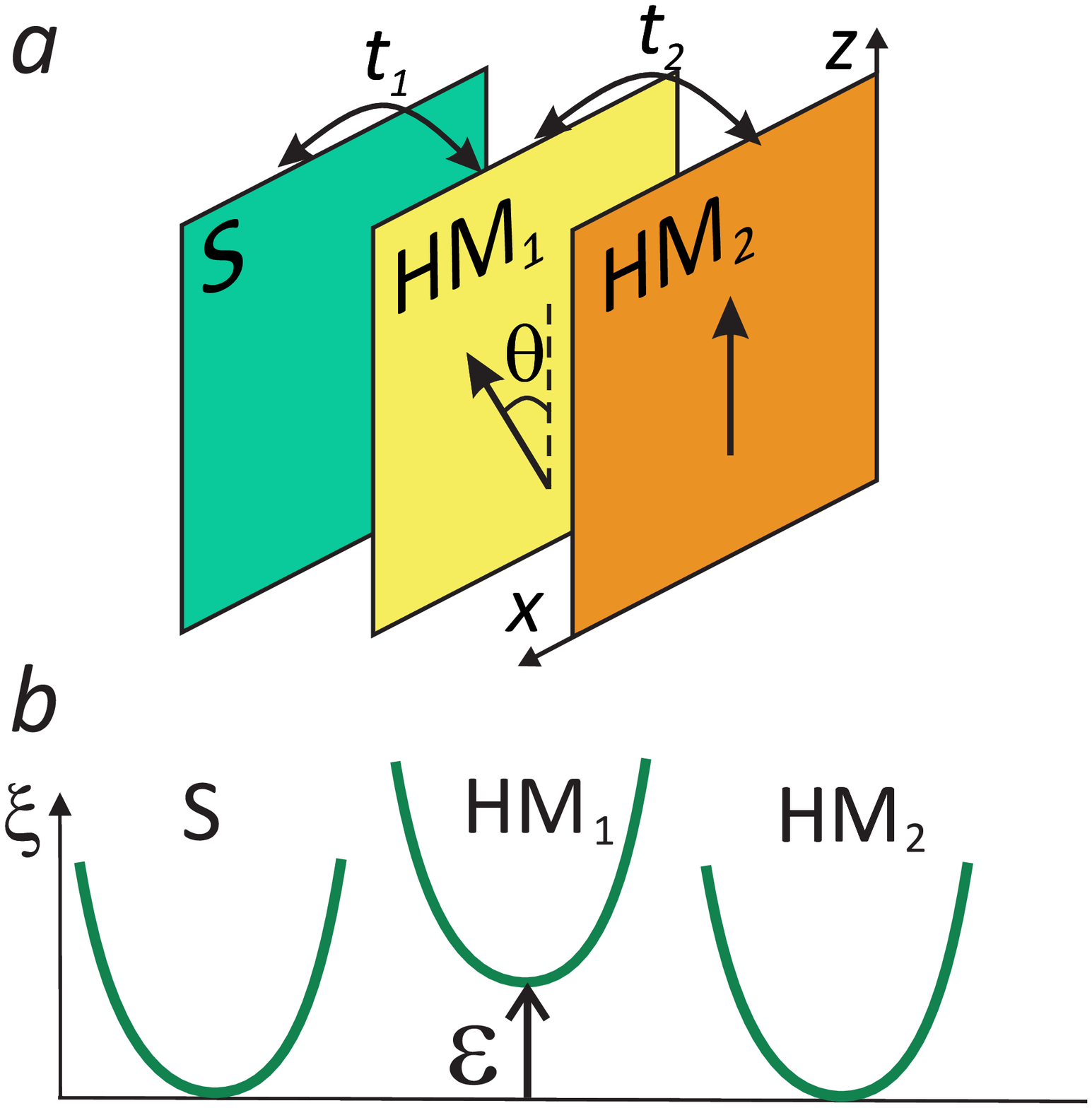}
\caption{(Color online) (a) Atomically thin S/HM$_1$/HM$_2$ spin valve.  The spin quantization axis in the central half-metal forms the angle $\theta$ with the $z$-axis, while the one in the HM$_2$ layer coincides with $z$-axis. The transfer integrals $t_1$ and $t_2$ couple the adjacent layers. (b) The electron energy band structure in each layer. The parameter $\varepsilon$ is the energy shift between the spin-up band of the HM$_1$ layer and the electron energy band in the superconductor.} \label{Fig1}
\end{figure}

In the present section we analyze the spin-valve effect in S/HM$_1$/HM$_2$ structures (see Fig.~\ref{Fig1}a) and calculate how the critical temperature $T_c$ depends on the angle $\theta$ between the spin quantization axes in two half-metals. The $y$-axis is chosen perpendicular to the layers interfaces. The spin quantization axis in the HM$_2$ layer is parallel to the $z$-axis, while the spin quantization axis in the HM$_1$ layer is assumed to lay in the $xz$-plane and form the angle $\theta$ with the $z$-axis. We assume that each layer has atomic thickness and the in-plane electron motion is ballistic. For simplicity we consider the limit of coherent electron tunneling between the layers which conserves the in-plane momentum. Moreover, the transfer integrals $t_1$ and $t_2$ coupling the superconductor with the HM$_1$ and HM$_2$ layer, respectively, are assumed to be much smaller than the superconducting critical temperature $T_c$. Such tight-binding model should be adequate for the description of the superconducting spin valves based, e.g., on La$_{0.7}$Ca$_{0.3}$MnO$_3$, a half-metallic compound \cite{Coey} which has been shown to have a significant effect on the properties of adjacent superconductor \cite{Nemes, Visani, Liu, Villegas}.  

To calculate the dependence $T_c(\theta)$ we use the Gor'kov formalism (see, e.g., Refs. \onlinecite{Tollis, Daumens1, Andreev, Prokic}). The system Hamiltonian consists of three terms:  
\begin{equation}
\label{H}
\hat H=\hat H_0+\hat H_S+\hat H_t.
\end{equation}
The first term
\begin{equation}
\label{H0}
\hat H_0=\sum \limits_{{\bf p};\alpha,\beta=\{1,2\}} \left[\xi({\bf p})\phi^+_{\alpha}\phi_{\beta} \delta_{\alpha \beta}+\hat V_{\alpha \beta}\psi^+_{\alpha}\psi_{\beta}+\hat W_{\alpha \beta}\eta^+_{\alpha}\eta_{\beta}\right]
\end{equation}
describes the quasiparticle motion in the normal state in each isolated layer, the second term
\begin{equation}
\label{Hs}
\hat H_S=\sum \limits_{{\bf p}}\left(\Delta^*  \phi_{{\bf p},2}\phi_{-{\bf p},1}+\Delta  \phi^+_{{\bf p},1}\phi^+_{-{\bf p},2}\right)
\end{equation}
describes the s-wave Cooper pairing in the S layer, and the last term
\begin{equation}
\label{Ht}
\hat H_t=\sum \limits_{{\bf p};\alpha=\{1,2\}}\left[t_1( \phi^+_{\alpha}\psi_{\alpha}+\psi^+_{\alpha}\phi_{\alpha})+t_2(\psi^+_{\alpha}\eta_{\alpha}+\eta^+_{\alpha}\psi_{\alpha})\right],
\end{equation}
characterizes the tunneling between the layers. In Eqs.~(\ref{H0})-(\ref{Ht}) $\phi$, $\psi$ and $\eta$ are the electron annihilation operators in the S, HM$_1$ and HM$_2$ layers, respectively, ${\bf p}$ is the quasiparticle momentum in the plane of the layers, $\xi({\bf p})$ is the electron energy spectrum in the S layer, $\alpha$ and $\beta$ are the spin indexes, $\Delta$ is the superconducting gap function. The matrices $\hat V$ and $\hat W$ describe the spin-dependent single-particle spectra in HM$_1$ and HM$_2$ layers, respectively. Let us denote $\xi_{\uparrow}$ and $\xi_{\downarrow}$ the energy spectra for the electrons with spin parallel (spin-up) and antiparallel (spin-down) to the spin quantization axis in the corresponding half-metallic layer. To take into consideration only the most important features of the band structure we assume that in HM$_2$ layer the energy spectrum for the spin-up electrons is the same as in the S layer $\xi_{\uparrow}=\xi({\bf p})$ while the spin-down energy band is not occupied ($\xi_{\downarrow}=+\infty$). The corresponding matrix $\hat{W}$ reads
\begin{equation}
\label{W}
\hat W=\left( \begin{array}{cc}
		\xi({\bf p}) & 0\\
		0 & \infty
	\end{array} \right).
\end{equation}
In the HM$_1$ layer it is convenient to assume the finite value of the exchange field $h$ and possible energy band shift $\xi_0$ with respect to the superconductor. This gives the matrix $\hat{V}$ in the form
\begin{equation}
\label{V}
\hat V=\left( \begin{array}{cc}
	\xi({\bf p})+\xi_{0}-h\cos \theta & -h\sin \theta\\
	-h\sin \theta & \xi({\bf p})+\xi_{0}+h\cos \theta
\end{array} \right),
\end{equation}
where $\theta$ is the angle between ${\bf h}$ and the $z$-axis. To approach the limit of the half-metal 
one should put simultaneously $h=+\infty$, $\xi_0=+\infty$ and $\xi_0-h=\varepsilon$. The resulting energy spectrum will contain only one band shifted by the value $\varepsilon$ with respect to the one in the superconductor (see Fig.~\ref{Fig1}b). 

In the limit of weak tunneling the critical temperature of the spin valve slightly differs from the critical temperature $T_{c0}$ of the isolated superconductor. Then it is convenient to represent the expression for $T_c(\theta)$ coming from the self-consistency equation in the following form:
\begin{equation}
\label{SCE}
T_c(\theta)=T_c(0)-T_{c0}^2 \sum \limits_{\omega_n=-\infty}^{+\infty} \int \limits_{\xi=-\infty}^{+\infty} d\xi \frac{\hat F^+_{12}(\theta)-\hat F^+_{12}(0)}{\Delta^*}.
\end{equation}
Here $\hat F^{+}_{\alpha \beta}=\langle T_{\tau}(\phi_{\alpha}^+,\phi_{\beta}^+)\rangle$ is the anomalous Green function in the superconductor, $T_{c0}$ is the critical temperature in the absence of the proximity effect ($t_1=t_2=0$), $\omega_n=\pi T_{c0}(2n+1)$ are the Matsubara frequencies. 

To calculate $\hat F^{+}$ we introduce the set of imaginary-time Green functions 
\begin{equation}
\label{GF1}
G_{\alpha,\beta}=-\langle T_{\tau}(\phi_{\alpha},\phi_{\beta}^+)\rangle, \qquad F^{+}_{\alpha,\beta}=\langle T_{\tau}(\phi_{\alpha}^+,\phi_{\beta}^+)\rangle,
\end{equation}
\begin{equation}
\label{GF2}
E^{\psi}_{\alpha,\beta}=-\langle T_{\tau}(\psi_{\alpha},\phi_{\beta}^+)\rangle, \qquad F^{\psi+}_{\alpha,\beta}=\langle T_{\tau}(\psi_{\alpha}^+,\phi_{\beta}^+)\rangle,
\end{equation}
\begin{equation}
\label{GF3}
E^{\eta}_{\alpha,\beta}=-\langle T_{\tau}(\eta_{\alpha},\phi_{\beta}^+)\rangle, \qquad F^{\eta+}_{\alpha,\beta}=\langle T_{\tau}(\eta_{\alpha}^+,\phi_{\beta}^+)\rangle.
\end{equation}
Next we obtain the system of Gor'kov equations taking the imaginary-time derivatives of above Green functions in the Fourier representation and using the Heisenberg equations for the operators $\phi$, $\psi$ and $\eta$:
\begin{gather}
\label{GS}
(i\omega_n-\xi)G+\Delta I F^{+}-t_1E^{\psi}=\hat 1, \\
(i\omega_n+\xi)F^+ -\Delta^* I G+t_1F^{\psi+}=0, \\
(i\omega_n-\hat V)E^{\psi}-t_1 G -t_2 E^{\eta}=0, \\
(i\omega_n+\hat V)F^{\psi+}+t_1 F^+ + t_2 F^{\eta+}=0, \\
(i\omega_n-\hat W)E^{\eta}-t_2 E^{\psi}=0, \\
(i\omega_n+\hat W)F^{\eta+}+t_2 F^{\psi+}=0.
\end{gather}

The above system enables an exact analytical solution for ${\hat F^+}$. In the first order  perturbation theory with the gap potential as a small parameter the result is
\begin{multline}
\label{AGF1}
\frac{\hat F^+}{\Delta^*}=\biggl\{(i\omega_n +\xi)\hat 1 -t_1^2\biggl[(i\omega_n +\hat V)-t_2^2(i\omega_n +\hat W)^{-1}\biggr]^{-1}\biggr\}^{-1} \\ \times \hat I \biggl\{(i\omega_n -\xi)\hat 1-t_1^2\biggl[(i\omega_n -\hat V)-t_2^2(i\omega_n -\hat W)^{-1}\biggr]^{-1} \biggr\}^{-1},
\end{multline}
where $\hat I=i \sigma_y$. 

The further substitution of (\ref{AGF1}) into (\ref{SCE}) gives the desired critical temperature. Since the transfer integrals are assumed to be small in comparison with $T_{c0}$ before substitution into (\ref{SCE}) we take the power expansion of (\ref{AGF1}) over $t_1$ and $t_2$ up to the forth order (see Appendix  \ref{App:AGF_SHM1HM2}) and obtain the explicit analytical result for $T_c$:
\begin{equation}
T_c(\theta)=T_c(0)+\sum \limits_{\omega_n=-\infty}^{+\infty} \int \limits_{-\infty}^{+\infty}\frac{T_{c0}^2t_1^2 t_2^2h(1-\cos \theta)(\omega_+  +\xi_0)d\xi}{\omega_+^3 \omega_- [(\omega_+ +\xi_0)^2-h^2]^2},
\end{equation}
where $\omega_{\pm}=i\omega_n \pm \xi$. Integrating over $\xi$ we find:
\begin{equation}
T_c(\theta)=T_c(0)-\sum \limits_{\omega_n>0}\Re\left\{ \frac{\pi T_{c0}^2t_1^2 t_2^2h(1-\cos \theta)(2i\omega_n+\xi_0)}{\omega_n^3[(2i\omega_n+\xi_0)^2-h^2]^2}\right\}.
\end{equation}

Finally, taking $h=+\infty$, $\xi_0=+\infty$ and $\xi_0-h=\varepsilon$ we obtain the critical temperature of S/HM$_1$/HM$_2$ spin valve in which the spin-up band in the central half-metal is shifted by the value $\varepsilon$ with respect to the energy band in the S layer:
\begin{equation}
\label{Tc_SHMHM}
T_c(\theta)=T_c(0)+\sum \limits_{\omega_n>0}\frac{\pi T_{c0}^2t_1^2 t_2^2(4\omega_n^2-\varepsilon^2)(1-\cos \theta)}{4\omega_n^3(4\omega_n^2+\varepsilon^2)^2}.
\end{equation}

\begin{figure}[t!]
	\includegraphics[width=0.34\textwidth]{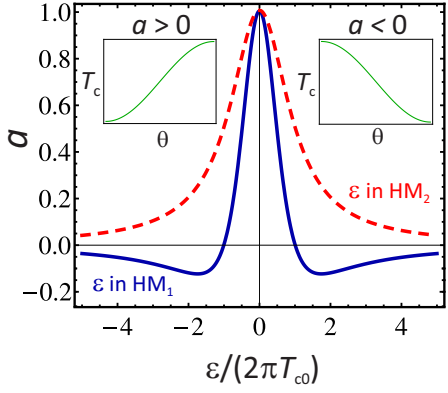}
	\caption{(Color online) The critical temperature is represented as $T_c(\theta)=T_c(0)+aT_{c0}[t/(2\pi T_{c0})]^4(1-\cos \theta)$. The dependences of the parameter $a$ on the energy shift $\varepsilon$ for the S/HM$_1$/HM$_2$ spin valve in which the spin-up band is shifted in the HM$_1$ half-metal (solid blue curve) or in the HM$_2$ layer (dashed red curve). Insets: the corresponding dependencies of the critical temperature on the angle $\theta$.} \label{Fig3}
\end{figure}

For the further analysis it is convenient to represent the expression for the critical temperature as
\begin{equation}
\label{Tc_gf}
T_c(\theta)=T_c(0)+a\frac{t_1^2t_2^2}{(2\pi T_{c0})^4}T_{c0}(1-\cos \theta).
\end{equation}
The sign of the parameter $a$ determines whether the standard ($a>0$) or inverse ($a<0$) spin-valve effect is realized in the system. Comparing Eq.~(\ref{Tc_gf}) with Eq.~(\ref{Tc_SHMHM}) we find:
\begin{equation}\label{a_1}
a=\sum\limits_{n\geq 0}\frac{(2n+1)^2-(\varepsilon/2\pi T_{c0})^2}{(2n+1)^3\left[(2n+1)^2+(\varepsilon/2\pi T_{c0})^2\right]^2} + O(t^2).
\end{equation}
If the occupied spin bands in both half-metals coincide with the electron energy band in the superconductor, i.e. $\varepsilon=0$, then $a>0$ and $T_c(\pi)$  is higher than $T_c(0)$ (standard spin-valve effect). However, if $\varepsilon \ne 0$ it is not always the case and the inverse switching is possible (see Fig.~\ref{Fig3} where we have put $t_1=t_2\equiv t$). Indeed, the coefficient $a$ becomes negative at $|\varepsilon|=\varepsilon_{cr} \sim 2\pi T_{c0}$ for $|\varepsilon|>\varepsilon_{cr}$ which corresponds to the monotonically decreasing  dependence $T_c(\theta)$. Note that for $\varepsilon \gg T_{c0}$ the coefficient $a$ can be estimated as $a \propto -(\pi T_{c0}/\varepsilon)^2$.

\begin{figure}[t!]
	\includegraphics[width=0.3\textwidth]{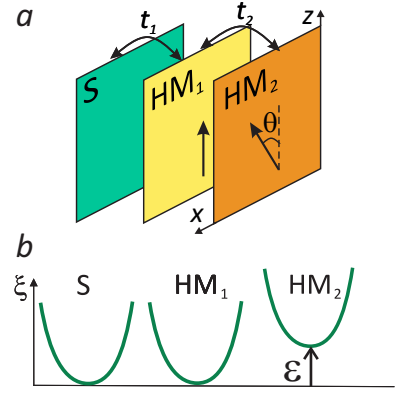}
	\caption{(Color online) The sketch (a) and the band structure (b) of the S/HM$_1$/HM$_2$ spin valve in which the spin-up band in the HM$_2$ half-metal is shifted by the value $\varepsilon$ with respect to the electron energy band in the S layer. The spin quantization axes in half-metals form the angle $\theta$ with each other.} \label{Fig2}
\end{figure}

Now we investigate if the inverse switching is possible in the case when the spin-up band is shifted in the HM$_2$ layer instead of HM$_1$ one. The corresponding band structure is shown in Fig.~\ref{Fig2}. For convenience, we assume that the $z$-axis coincides with the spin quantization axis in the HM$_1$ layer and forms the angle $\theta$ with one in the HM$_2$ half-metal (see Fig.\ref{Fig2}). The Hamiltonian, the system of Gor'kov equations and its solution for the anomalous Green function still have the form (\ref{H})--(\ref{Ht}), (\ref{GS}) and (\ref{AGF1}), respectively, if one replaces $\hat V \rightarrow \hat W$ and $\hat W \rightarrow \hat V$. Substituting the expansion of the Green function $\hat F^{+}$ over $t_1$ and $t_2$ into (\ref{SCE}) we obtain the critical temperature up to the forth order over the transfer integrals:
\begin{equation}
T_c(\theta)=T_c(0)+\sum \limits_{\omega_n=-\infty}^{+\infty} \int \limits_{-\infty}^{+\infty}\frac{T_{c0}^2t_1^2 t_2^2h(1-\cos \theta)d\xi}{\omega_+^4 \omega_- [(\omega_+ +\xi_0)^2-h^2]}.
\end{equation}
Next we integrate over $\xi$ and find
\begin{equation}
T_c(\theta)=T_c(0)-\sum \limits_{\omega_n>0} \frac{\pi T_{c0}^2t_1^2 t_2^2h\xi_0(1-\cos \theta)}{[4\omega_n^2+(h-\xi_0)^2][4\omega_n^2+(h+\xi_0)^2]}.
\end{equation}
Finally, taking $h=+\infty$, $\xi_0=+\infty$ and $\xi_0-h=\varepsilon$ we obtain the critical temperature of S/HM$_1$/HM$_2$ spin valve:
\begin{equation}\label{Tcres}
T_c(\theta)=T_c(0)+\sum \limits_{\omega_n>0}\frac{\pi T_{c0}^2t_1^2 t_2^2(1-\cos \theta)}{4\omega_n^3(4\omega_n^2+\varepsilon^2)}.
\end{equation}
The corresponding parameter $a$ reads as:
\begin{equation}\label{a_2}
a=\sum\limits_{n\geq 0}\frac{1}{(2n+1)^3\left[(2n+1)^2+(\varepsilon/2\pi T_{c0})^2\right]} +O(t^2).
\end{equation}
From Eq.~(\ref{Tcres}) one sees that  $T_c(\pi)>T_c(0)$ for any $\varepsilon$ (see Fig.~\ref{Fig3}). Thus, the shift of the occupied spin band in the side half-metal does not give rise to the inverse spin-valve effect.

\section{Spin valve effect in HM1/S/HM2 structures} \label{Sec:HM1SHM2}

\begin{figure}[t!]
	\includegraphics[width=0.3\textwidth]{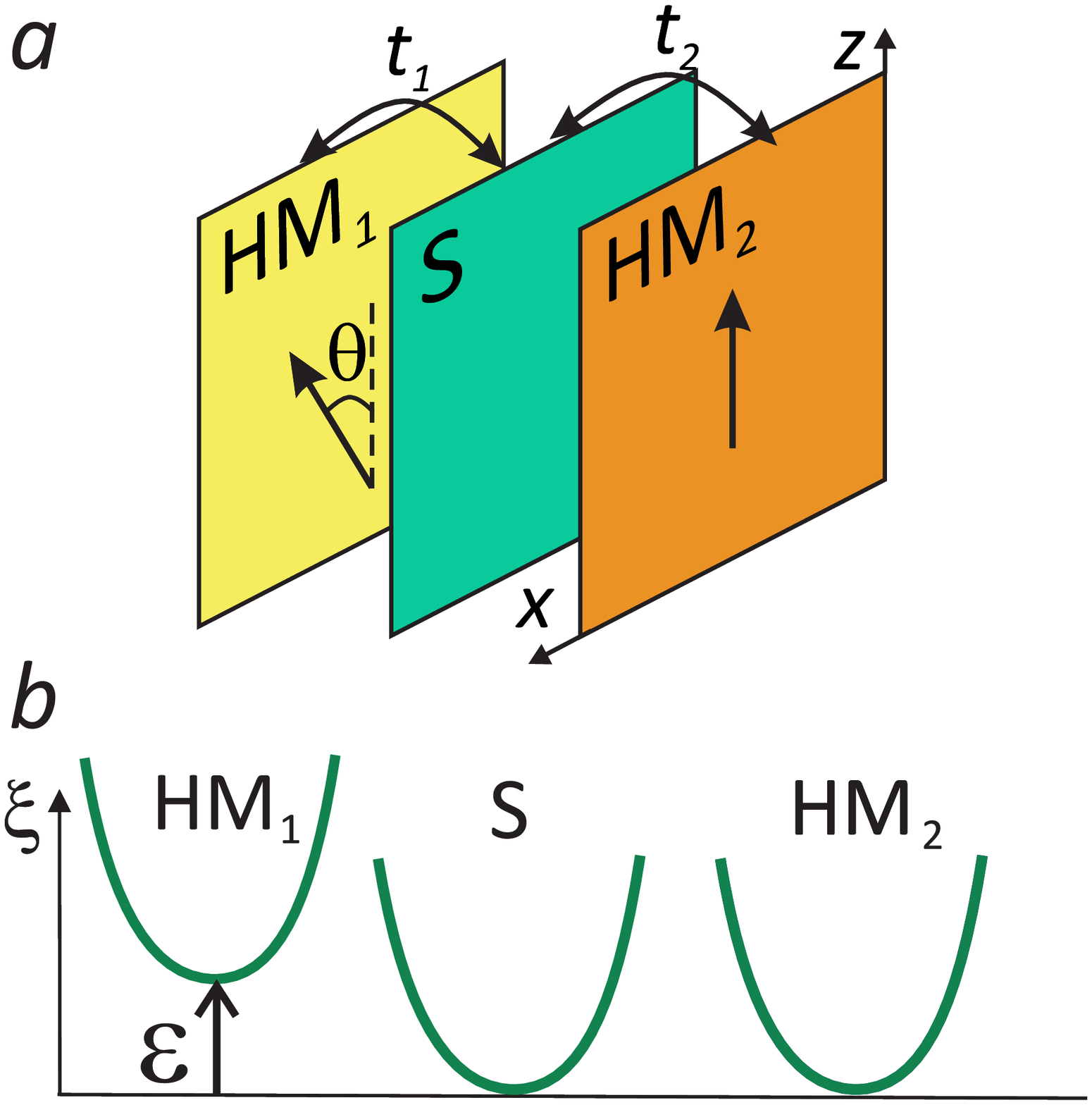}
	\caption{(Color online) The HM$_1$/S/HM$_2$ structure of atomic thickness. (a) The sketch of the spin valve. Here $\theta$ is the angle between spin quantization axes in the HM$_1$ and HM$_2$ layers. The HM$_1$ half-metal and the superconductor are coupled by the transfer integral $t_1$, while the transfer integral $t_2$ couples S and HM$_2$ layers. (b) The band structure of the spin valve.} \label{Fig4}
\end{figure}

In this section we consider spin valves which consist of a superconductor placed between two half-metals (see Fig.\ref{Fig4}) and calculate the critical temperature of such structure. The spin quantization axis in the HM$_2$ layer is directed along the $z$-axis and forms the angle $\theta$ with the one in the HM$_1$ layer. The spin-up band in the right half-metal coincides with the energy band in the superconductor while in the left half-metal we assume $\xi_{\uparrow}=\xi({\bf p})+\varepsilon$. The Hamiltonian has the form (\ref{H}) with $H_0$, $H_s$, $\hat W$ and $\hat V$ satisfying (\ref{H0}), (\ref{Hs}), (\ref{W}), and (\ref{V}), respectively, and
\begin{equation}
\label{Ht2}
\hat H_t=\sum \limits_{{\bf p};\alpha=\{1,2\}}\left[t_1( \phi^+_{\alpha}\psi_{\alpha}+\psi^+_{\alpha}\phi_{\alpha})+t_2(\phi^+_{\alpha}\eta_{\alpha}+\eta^+_{\alpha}\phi_{\alpha})\right].
\end{equation}

\begin{figure}[b!]
	\includegraphics[width=0.35\textwidth]{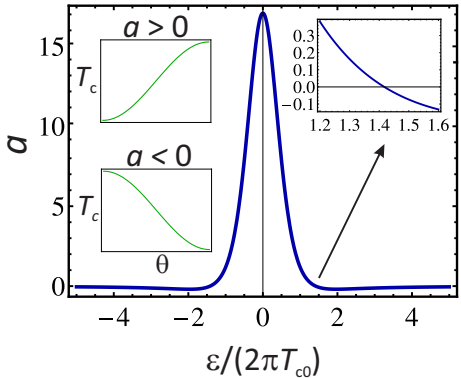}
	\caption{(Color online) The critical temperature has the form $T_c(\theta)=T_c(0)+aT_{c0}[t/(2\pi T_{c0})]^4(1-\cos \theta)$. The parameter $a$ vs. the energy separation $\varepsilon$ for the HM$_1$/S/HM$_2$ structure. The left insets demonstrate corresponding dependencies of the critical temperature $T_c$ vs. the angle $\theta$. The right inset shows in details the part of the main plot where $a$ changes the sign. } \label{Fig5}
\end{figure}

As before, we introduce the minimal set of the Green functions (\ref{GF1})--(\ref{GF3}) required for the calculation of $T_c$ and write down the system of Gor'kov equations (see Appendix \ref{App:HM1SHM2}). Their solution for the anomalous Green function in the linear approximation over the gap potential reads:
\begin{multline}
\label{AGF2}
\frac{\hat F^+}{\Delta^*}=\left[(i\omega_n +\xi)\hat 1 -t_1^2(i\omega_n +\hat V)^{-1}-t_2^2(i\omega_n +\hat W)^{-1}\right]^{-1} \times \\ \times \hat I \left[(i\omega_n -\xi)\hat 1-t_1^2(i\omega_n -\hat V)^{-1}-t_2^2(i\omega_n -\hat W)^{-1}\right]^{-1}.
\end{multline}
Expanding the expression (\ref{AGF2}) over $t_1$ and $t_2$ up to forth order and substituting it into (\ref{SCE}) we find:
\begin{multline}
T_c(\theta)=T_c(0)+\sum \limits_
{\omega_n=-\infty}^{+\infty} \int \limits_{-\infty}^{+\infty}\frac{T_{c0}^2t_1^2 t_2^2h(1-\cos \theta)d\xi}{\omega_+^3 \omega_-}\times \\ \times \left\{\frac{2}{\omega_+[(\omega_++\xi_0)^2-h^2]}+\frac{1}{\omega_-[(\omega_--\xi_0)^2-h^2]}\right\}.
\end{multline}

Performing the same analysis as before, we obtain the critical temperature of HM$_1$/S/HM$_1$ spin valve of atomic thickness with the spin-up band in HM$_1$ shifted by the value $\varepsilon$ with respect to the electron energy band in the superconductor up to the forth order over the transfer integrals:
\begin{equation}
\begin{array}{l}{\ds T_c(\theta)=T_c(0)+}\\{}\\{\ds+\sum \limits_{\omega_n>0}\frac{\pi T_{c0}^2t_1^2 t_2^2(68\omega_n^4-7\varepsilon^2 \omega_n^2-\varepsilon^4)(1-\cos \theta)}{4\omega_n^3(4\omega_n^2+\varepsilon^2)^3}.}\end{array}
\end{equation}
In the limit $t_1 \rightarrow 0, t_2 \rightarrow 0$ the corresponding parameter $a$ reads as:
\begin{equation}\label{a_3}
a=\sum\limits_{n\geq 0}\frac{17(2n+1)^4-7(2n+1)^2(\varepsilon/2\pi T_{c0})^2-(\varepsilon/2\pi T_{c0})^4}{4(2n+1)^3\left[(2n+1)^2+(\varepsilon/2\pi T_{c0})^2\right]^3}.
\end{equation}
One sees that the behavior $T_c(\theta)$ strongly depends on the value of the energy shift $\varepsilon$ (see Fig.~\ref{Fig5}). Since the parameter $a$ changes the sign at $|\varepsilon|=\varepsilon_{cr} \sim 2.8\pi T_{c0}$ the system reveals the standard spin-valve effect for $|\varepsilon|< \varepsilon_{cr}$ and the inverse one in the opposite case.

Note that the sign change of the difference $[T_c(\pi)-T_c(0)]$ is confirmed by the numerical solution of the self-consistency equation with the exact anomalous Green function Eq.(\ref{AGF2}), see Appendix \ref{App:Exact}. Thus, this result is not the matter of the approximation (the expansion over $t_1$ and $t_2$).

\section{Conclusion} \label{Sec:Conclusion}

We developed the theory of the spin-valve effect in the atomically thin S/HM$_1$/HM$_2$ and HM$_1$/S/HM$_2$ structures beyond the quasiclassical approximation.  We show that the details of the electron energy band structure in half-metallic layers strongly affect the behavior of system critical temperature $T_c$: depending on the position of the only occupied spin band in one of the HM$_1$ layers the dependence of $T_c$ on the angle between the spin quantization axes in half-metals can be either monotonically increasing or decreasing which corresponds to the standard or inverse spin-valve effect, respectively.  

The recent experiments on the spin valves containing the half-metallic La$_{0.7}$Ca$_{0.3}$MnO$_3$ compound demonstrated that this strongly-polarized ferromagnet is more stable compared to CrO$_2$\cite{Robinson} and gives rise to the anomalous behavior of $T_c$.\cite{Nemes, Visani, Liu, Villegas} We hope that such stability will allow to fabricate complex spin valves with two HM layers and perform experimental verification of our results. Note that our model does not account the finite thickness of the HM layers. If this thickness is much smaller than the superconducting correlation length $\xi_h$ inside the half-metal our results should remain qualitatively the same. At the same time, for the HM layers of the thickness $\sim \xi_h$ the interference effects may have a significant impact on $T_c$. As the result, we expect that the type of the spin valve effect will be determined by the combination of two factors: the influence of the band structure details and the interference effects.

\acknowledgements

The authors thank A. Buzdin and A. S. Mel'nikov for stimulating discussions. This work was supported by Russian Science Foundation under Grant No. 15-12-10020 (SM), Russian Presidential Scholarship SP3938.2018.5 (SM), and Foundation for the advancement of theoretical physics "BASIS" (ZhD and SM). The work of ZhD was carried out within the framework of the state task.

\appendix

\section{Anomalous Green function in S/HM1/HM2 spin valve of atomic thickness}\label{App:AGF_SHM1HM2}

Expanding Eq.(\ref{AGF1}) over $t_1$ and $t_2$ up to the fourth order we obtain the following expression:

\begin{multline}
	\frac{\hat F^{+}}{\Delta^*} \simeq \frac{1}{\omega_+ \omega_-}\biggl[\hat I +t_1^2\biggl(\frac{1}{\omega_+}\hat X_+\hat I+\frac{1}{\omega_-}\hat I\hat X_-\biggr) + \\+ t_1^2t_2^2\biggl(\frac{1}{\omega_+}\hat X_+\hat Y_+ \hat X_+\hat I+\frac{1}{\omega_-}\hat I\hat X_-\hat Y_- \hat X_-\biggr) +\\+t_1^4\biggl(\frac{1}{\omega_+^2}\hat X_+^2\hat I+\frac{1}{\omega_-^2}\hat I\hat X_-^2\biggr)\biggr],
\end{multline}

where $\hat X_{\pm}=\left(i\omega_n \hat 1 \pm \hat V \right)^{-1}$, $\hat Y_{\pm}=\left(i\omega_n \hat 1 \pm \hat W \right)^{-1}$.

	\section{Gor'kov equations for the HM1/S/HM2 structure} \label{App:HM1SHM2}
	
	Using the same procedure as before, we obtain the following system of Gor'kov equations
	\begin{gather*}
	(i\omega_n-\xi)G+\Delta I F^{+}-t_1E^{\psi}-t_2 E^{\eta}=\hat 1, \\
	(i\omega_n+\xi)F^+ -\Delta^* I G+t_1F^{\psi+}+t_2F^{\eta+}=0, \\
	(i\omega_n-\hat V)E^{\psi}-t_1 G=0, \\
	(i\omega_n+\hat V)F^{\psi+}+t_1 F^+ =0, \\
	(i\omega_n-\hat W)E^{\eta}-t_2 G=0, \\
	(i\omega_n+\hat W)F^{\eta+}+t_2 F^{+}=0.
	\end{gather*}
	
	\section{Exact results for critical temperature of S/HM1/HM2 spin valve of atomic thickness}\label{App:Exact}
\begin{figure}[b!]
		\includegraphics[width=0.35\textwidth]{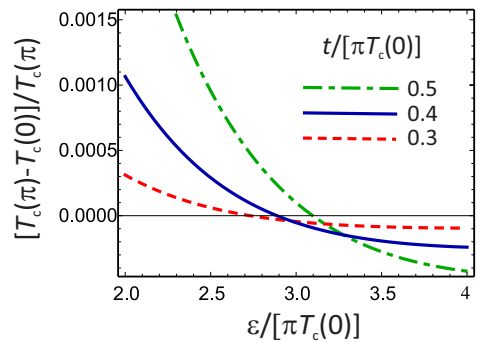}
		\caption{(Color online) The exact results for the critical temperature vs. the energy separation $\varepsilon$ of the HM$_1$/S/HM$_2$ structure.} \label{Fig6}
	\end{figure}

	Our model enables to obtain the exact solution for the critical temperature, which is valid at all $t_1,t_2 \lesssim \Delta$. Solving numerically the self-consistency equation with the exact anomalous Green function (\ref{AGF2}), we obtain the sign change of the difference $[T_c(\pi)-T_c(0)]$ at nonzero $\epsilon$ even for not very small transfer integrals $t_1,t_2 \lesssim \pi T_{c0}$ [see Fig.(\ref{Fig6})]. This confirms the sign change of the coefficient $a$, obtained analytically in the limit $t_1,t_2 \rightarrow 0$ [see Eq.(\ref{a_3}) and Fig.(\ref{Fig5})].

\end{document}